\documentclass{cs21proc}

\usepackage{kantlipsum}

\editors{A. S. Brun, J. Bouvier, P. Petit}
\publisher{Zenodo}
\conference{The 21th Cambridge Workshop on Cool Stars, Stellar Systems, and the Sun}
\conferencedate{2022}

\title{Surface activity of the G dwarf primary in the quaternary star system V815\,Her}
\author{Zsolt Kővári,$^{1,2}$
        Klaus G. Strassmeier,$^{3}$
        Tamás Borkovits,$^{1,4,5}$
        Levente Kriskovics,$^{1,2}$
        Katalin Oláh,$^{1,2}$
        Bálint Seli,$^{1,2}$
        Krisztián Vida,$^{1,2}$}

\affiliation{$^{1}$ Konkoly Observatory, Research Centre for Astronomy and Earth Sciences, Budapest, Hungary\\
        $^{2}$ CSFK, MTA Centre of Excellence\\
        $^{3}$ Leibniz-Institute for Astrophysics, Potsdam, Germany\\
        $^{4}$ Baja Astronomical Observatory of University of Szeged, Baja, Hungary\\ 
        $^{5}$ ELKH--SZTE Stellar Astrophysics Group}

\shorttitle{Surface activity of V815\,Her}
\shortauthors{Zs. Kővári et al.}

\abs{We investigate the magnetic activity of the G dwarf primary star in the multiple system V815\,Herculis. Recently, TESS Sector 26 data have revealed that V815\,Her is in fact a four-star system consisting of two close binaries in a long-period orbit. We give preliminary orbital solution for the long-known but unseen "third body" V815\,Her~`B', which is itself a close eclipsing binary of two M dwarfs. Long-term spot activity of the G dwarf is presented along with the very first Doppler image reconstructions of its spotted surface.}

\begin{document}

\maketitle

\section{V815\,Her: A multiple system in the solar neighborhood}

Close binaries that contain magnetically active component may be regarded as astrophysical laboratory for studying the effect of binarity on activity. Especially interesting are those binaries and multiple star systems which contain active G-type main-sequence (MS) stars with inner structure (convective zone) similar to that of the Sun, making it possible to investigate how the solar-type dynamo would work under different circumstances. Accordingly, in the following study, we investigate the magnetic activity of the G primary star in the quaternary star system V815\,Her (=HD\,166181), which is 32~pc from the Sun.

The primary G-type dwarf star together with an unseen, but very likely M-dwarf companion form the $P_{\rm orb}=1.8$\,d orbital period close binary system V815\,Her~`A'. It has long been known that the active G (most probably G5-6~V) component V815\,Her~`Aa' rotates synchronously with the orbital period and features cool spots on its surface \citep{1980IBVS.1791....1M}. Moreover, the strong Li line in the optical spectrum \citep{2005AJ....129.1001F} indicates that the G star is young, near the zero-age main sequence (ZAMS). This observation is consistent with the star being listed in the Two Micron All Sky Survey Point Source Catalog \citep{2009ApJS..184..138H} due to its infrared excess, most probably related to primordial protoplanetary material.

\citet{2004RMxAC..21...45F} calculated long-period orbital elements for the outer companion V815\,Her~`B' for the first time with a preliminary period of 6.3\,yr. With additional radial velocity measurements from 2003 this value was refined to 5.73\,yr \citep{2005AJ....129.1001F}. From their mass estimates for the system components of the long-period orbit \citet{2005AJ....129.1001F} also suggested that the third (still unseen) component might also be a close binary.

\section{Long-term photometric behavior}

\begin{figure}[hb]
	\centering
	\includegraphics[width=\linewidth]{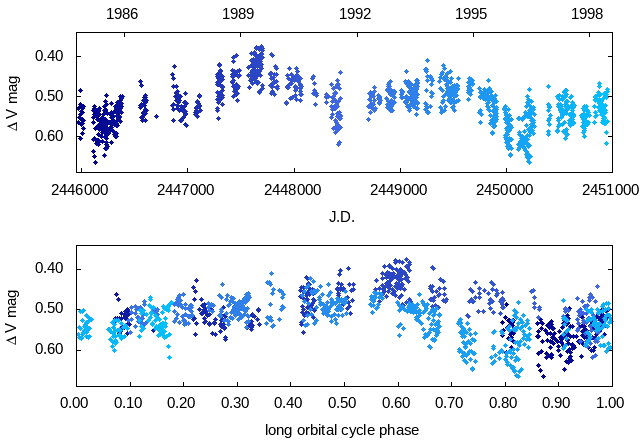}
	\caption{{\it Top}: APT $V$ light curve of V815\,Her between 1984-98. {\it Bottom}: Folded light curve using the 2092.2\,d period of the wide orbit. As zero phase the time of the periastron passage was set. The continuous color gradient helps to distinguish successive phases.}
	\label{fig:fig_wide}
\end{figure}

In the top panel of Fig.~\ref{fig:fig_wide} we plot the $V$ observations of V815\,Her between 1984-98 from \cite{2000A&A...362..223J}. In these data we found a long-term brightness modulation, which showed a very good agreement with the approximate period of $\approx$2100\,d of the wide orbit. In the bottom panel of Fig.~\ref{fig:fig_wide} we plot the $V$ data along with the folded light curve phased with the period of the long orbit taken from \cite{2005AJ....129.1001F}. 
At this point we speculate that this modulation may be related in some way to the presence (precession?) of a protoplanetary disk.

\section{New orbital solution for V815\,Her~Ba+Bb}
\begin{figure}[t]
	\centering
	\includegraphics[width=\linewidth]{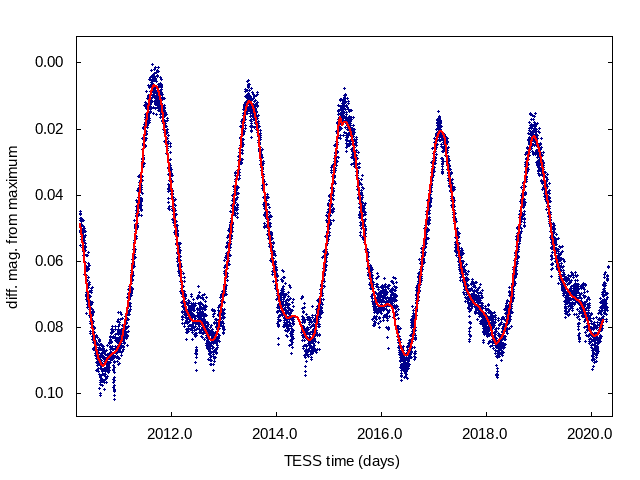} \includegraphics[width=\linewidth]{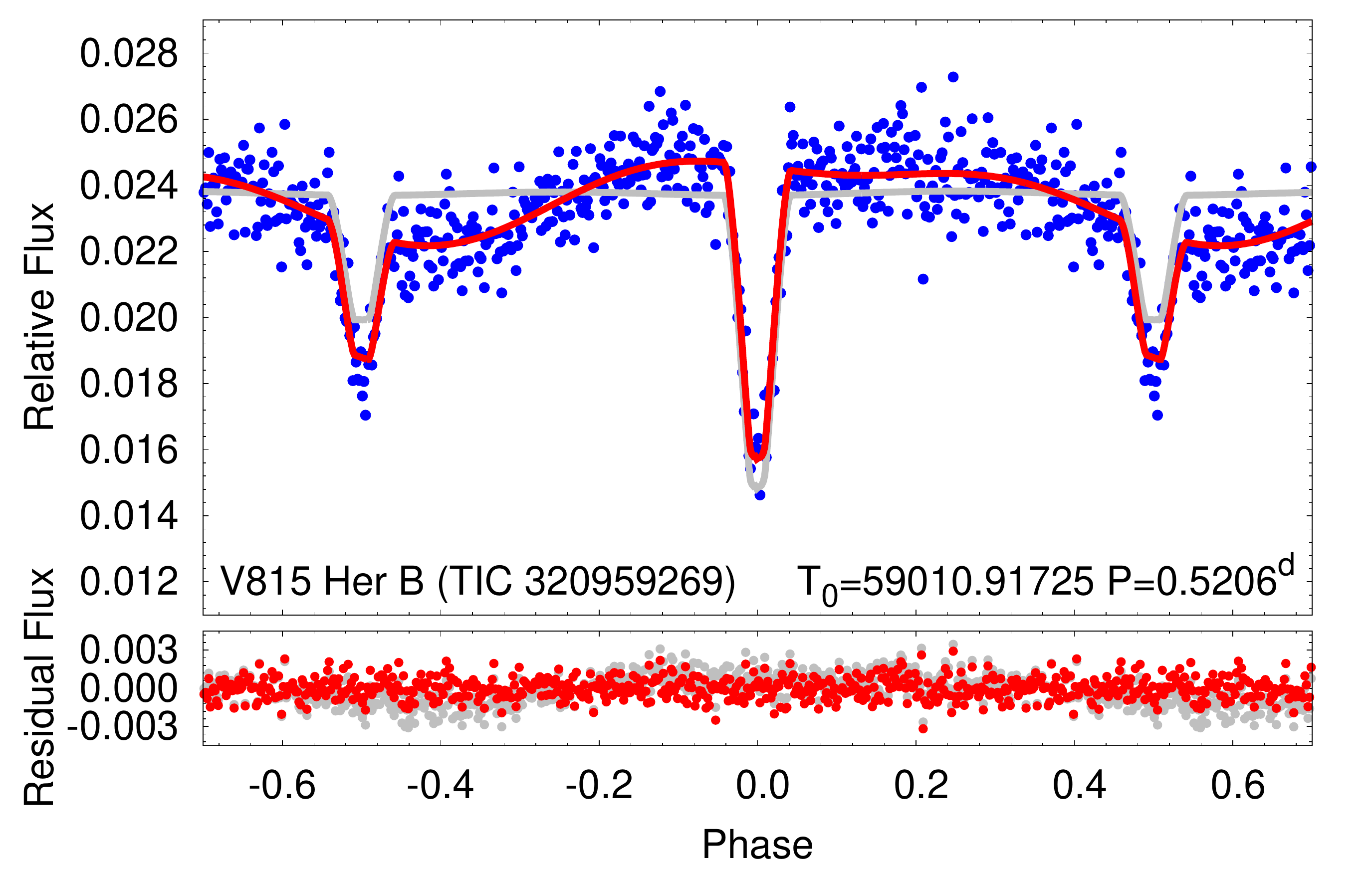}
	\caption{{Top:} \emph{TESS} Sector 26 light curve of V815\,Her (blue dots). The 1.8-d rotational modulation is fitted with a time-series three-spot model (red line). {Bottom:} Cleaned and folded \emph{TESS} light curve (blue dots) for the wide binary companion V815\,Her~`B'. The original orbital solution without assuming surface inhomogeneities is drawn with grey line, while the model drawn with red line fits also the light curve variations arising from stellar spots. Below are the residuals.}
	\label{fig:TESS}
\end{figure}

V815\,Her (=TIC~320959269) is listed as a \emph{TESS} planet host candidate with a supposed planetary orbit of 0.26\,d. So far, \emph{TESS} has observed V815\,Her during Sector 26, 40, and 53.
However, after removing the spot variability of the G star using the 1.8\,d rotational period (see Fig.~\ref{fig:TESS}, top), from a closer inspection it became clear to us that the depths of successive minima are slightly different, i.e., primary and secondary minima alternate. Evidently, instead a planet orbiting with a period of 0.26\,d we found a close eclipsing binary with a double-length period of 0.52\,d bound to V815\,Her~`A' in the outer orbit. The eclipsing binary light curve analysis for V815\,Her~`B' was carried out with the software package {\sc Lightcurvefactory} \citep[see][and further references therein]{2019MNRAS.483.1934B,2020MNRAS.493.5005B}. The resulting model fit for the \emph{TESS} Sector 26 data is plotted in the bottom panel of Fig.\ref{fig:TESS} and the parameters of the orbital solution are given in Table~\ref{tab:T1}.

\section{Doppler images and surface evolution of the G-dwarf component}
\begin{figure}[t!!]
	\centering
	\includegraphics[width=\linewidth]{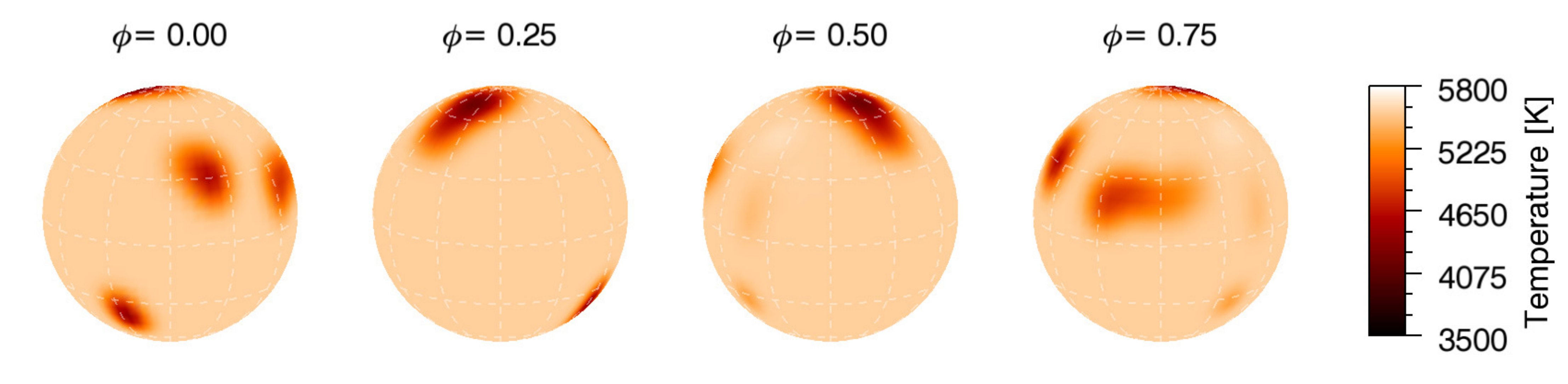}
    \includegraphics[width=\linewidth]{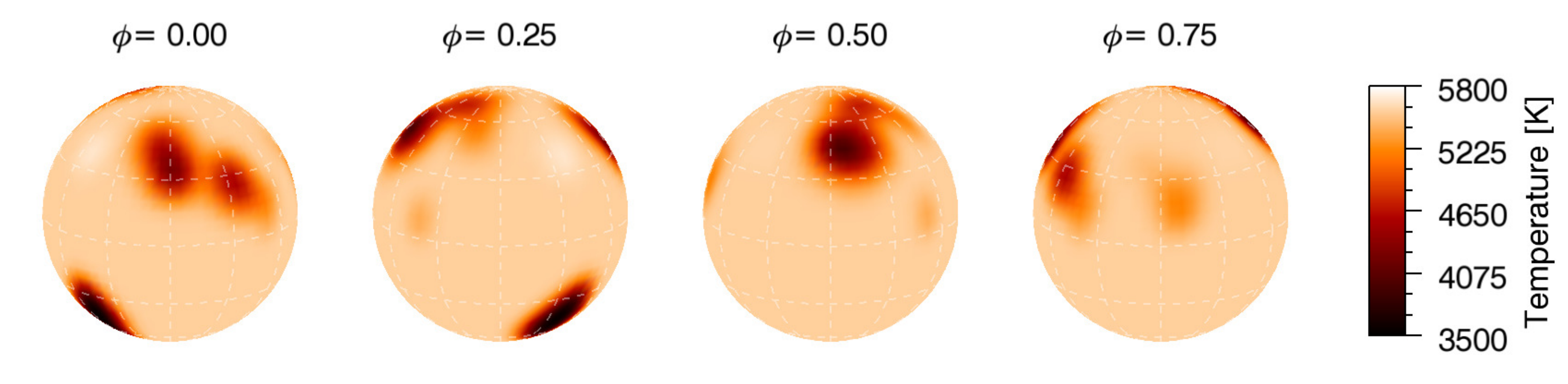}
	\caption{Two Doppler images of V815\,Her~Aa from 2018. The respective mid HJDs of the images are 2458274.624 (top) and 2458280.067 (bottom). Note the significant change in the spot distribution during the $\approx$3$P_{\rm rot}$ time interval, which may partially reflect surface differential rotation.}
	\label{fig:DI}
\end{figure}

Here we show preliminary results of a Doppler imaging study for the G star for the very first time. We used high-resolution optical spectra from the STELLA-II telescope of the STELLA robotic observatory taken during the 2018 observing run. For the inversion we used the \emph{iMap} code \citep{imap} with adopting  $i=75^{\circ}$ inclination of the rotational axis and $v\sin i$ of 30\,kms$^{-1}$ \citep{2005AJ....129.1001F}. In Fig.~\ref{fig:DI} two surface spot reconstructions are plotted as examples. The two consecutive maps reflect significant spot activity and spot distribution change on a time scale comparable to rotation (cf. Fig.~\ref{fig:TESS}, top). Some of the changes are probably due to differential rotation, which we will report in more detail in our forthcoming paper (Kővári {\it et al.,} 2023).

\begin{table*}[t]
	\centering
	\caption{Model parameters for V815\,Her~Ba+Bb}
	\label{tab:T1}
\begin{tabular*}{\linewidth}{l @{\extracolsep{\fill}} lll}
	\noalign{\smallskip}\hline\hline\noalign{\smallskip}
\multicolumn{3}{c}{Orbital elements} \\
	\noalign{\smallskip}\hline\noalign{\smallskip}
  $P$ [days] & \multicolumn{2}{c}{$0.5206$}  \\
  $a$ [R$_\odot$] & \multicolumn{2}{c}{$2.33_{-0.24}^{+0.29}$}\\
  $e$ & \multicolumn{2}{c}{$0$}\\
  $\omega$ [deg] & \multicolumn{2}{c}{$-$} \\ 
  $i$ [deg] & \multicolumn{2}{c}{$89.17_{-1.45}^{+0.69}$}  \\
  $T_0$ [BJD - 2400000] & \multicolumn{2}{c}{$59011.04717_{-0.0002}^{+0.0002}$}  \\
  mass ratio $[q=m_\mathrm{sec}/m_\mathrm{pri}]$ & \multicolumn{2}{c}{$0.48_{-0.08}^{+0.12}$} \\
  	\noalign{\smallskip}\hline\noalign{\smallskip}
\multicolumn{3}{c}{Stellar parameters} \\
	\noalign{\smallskip}\hline\noalign{\smallskip}
   & Ba & Bb \\
  	\noalign{\smallskip}\hline\noalign{\smallskip}
 fractional radius [$R/a$] & $0.1603_{-0.0057}^{+0.0046}$ & $0.1102_{-0.0060}^{+0.0053}$ \\
 temperature relative to $(T_\mathrm{eff})_\mathrm{Ba}$ & $1$ & $0.8736_{-0.0293}^{+0.0222}$ \\
 fractional flux [in \textit{TESS}-band] & $0.0179_{-0.0021}^{+0.0020}$ & $0.0038_{-0.0003}^{+0.0003}$ \\
 $m$ [$M_\odot$] & $0.444_{-0.158}^{+0.198}$ & $0.185_{-0.022}^{+0.053}$ \\
 $R$ [$R_\odot$] & $0.377_{-0.051}^{+0.047}$ & $0.256_{-0.023}^{+0.026}$ \\
 $T_\mathrm{eff}$ [K] & $3702_{-174}^{+369}$ & $3245_{-91}^{+174}$ \\
 $L_\mathrm{bol}$ [$L_\odot$] & $0.024_{-0.010}^{+0.019}$ & $0.007_{-0.002}^{+0.003}$ \\
 $M_\mathrm{bol}$ & $8.77_{-0.64}^{+0.54}$ & $10.20_{-0.41}^{+0.34}$ \\
 $M_V           $ & $10.47_{-1.31}^{+1.04}$ & $13.20_{-0.41}^{+0.34}$ \\
 $\log g$ [dex] & $4.94_{-0.08}^{+0.05}$ & $4.91_{-0.06}^{+0.07}$ \\
 	\noalign{\smallskip}\hline\noalign{\smallskip}
\end{tabular*}
\end{table*}

\section*{Acknowledgments}
{The authors acknowledge the Hungarian National Research, Development and Innovation Office
grants OTKA K-131508, KKP-143986 (\'Elvonal), and 2019-2.1.11-T\'eT-2019-00056. LK acknowledges the Hungarian National Research, Development and Innovation Office
grant OTKA PD-134784. LK and KV are supported by the Bolyai J\'anos Research Fellowship of the Hungarian Academy of Sciences.
KV is supported by the Bolyai+ grant \'UNKP-22-5-ELTE-1093, BS is supported by the \'UNKP-22-3 New National Excellence Program of the Ministry for Culture and
Innovation from the source of the National Research, Development and Innovation Fund.
}

\bibliographystyle{cs21proc}
\bibliography{refs.bib}

\end{document}